# Magnet R&D for the Muon Collider

L. Bottura, B. Auchmann, F. Boattini, B. Bordini, B. Caiffi, L. Cooley, S. Fabbri, S. Gourlay, S. Mariotto, T. Nakamoto, S. Prestemon, M. Statera

**Summary**

A proton-driven Muon Collider, in the configuration that has resulted from the efforts of the International Muon Collider Collaboration (IMCC), poses multiple and exceptional magnet system challenges. Addressing these challenges will require a focused effort to advance accelerator magnet technology well beyond the present state of the art, including activities that have not previously been supported by High Energy Physics (HEP) programs, but are synergic with them. This proposal presents the motivation for a directed effort focusing on the development and testing of small- and full-scale magnet prototypes, ultimately culminating in their validation under collider-relevant conditions. This document summarizes technology status, challenges, and development targets, and outlines a detailed plan with staged milestones to advance the technological readiness of magnet systems, bringing the realization of the Muon Collider closer to reality. The total resources to achieve this goal are estimated at 82.5 MCHF and 414 FTE y over ten years, of which 39 MCHF and 199 FTE y are engaged over the first five years. Reaching the desired performance with sustainable technology will depend greatly on exploiting the potential of High Temperature Superconductors (HTS). Mainly because of this, the R&D proposed here has significant potential to broadly impact HEP and its other circular collider considerations such as the FCC-hh, as well as other fields of scientific and societal application, e.g. science in high magnetic fields, NMR and MRI, fusion and other power and mobility applications.

**Introduction**

The Muon Collider (MC) [1, 2] embodies a groundbreaking concept in circular colliders for high-energy physics, offering a unique pathway to achieve unprecedented energy, luminosity and efficiency (cost per parton collision), while significantly reducing footprint and environmental impact compared to conventional collider technologies. A critical aspect of its feasibility lies in the development of cutting-edge superconducting magnet systems capable of meeting the demanding requirements of muon production, acceleration, and collision. This was already recognized in the seminal US-based Muon Accelerator Program (US-MAP) study [3, 4] which produced an initial configuration that was crucial to identify the main challenges. In recent years, the International Muon Collider Collaboration (IMCC) [5], hosted at CERN, has evolved significantly this initial concept. The integrated design effort of IMCC was conducive to significant progress in both the conceptual and, in some cases, the engineering design of the magnet systems [6, 7]. The status of magnet development and other main findings can be found in references [8-22], and in the extended summary of [23] where we have reviewed performance requirements and identified the main challenges associated with the concept baseline.

In this proposal we recall the development targets for the Muon Collider magnets and highlight the associated challenges that are the main development drivers. We then briefly review the state of the art in solenoid, dipole and quadrupole magnets built with LTS and HTS technology, and we conclude with a gap analysis to identify and motivate the priority directions for future R&D. We then detail the R&D proposal, built to foster magnet technology for a muon collider. The majority of the targets of the R&D proposed are well beyond the present state of the art and are not pursued by other R&D or studies in HEP.

The proposed activity is based on the evaluation of the *Technology Readiness Level* (TRL) [29] and organized as a series of *Technology Milestones* (TM). Each TM is a package of development activities that leads to a small- and full-size magnet or system demonstrator, carefully selected to coalesce efforts around practical realizations. Each demonstrator is intended to be both a direct proof of performance, as well as a driver for the development of supporting magnet science and technology. The proposal focuses on a time span of ten years, and the values of material and personnel resources quoted for each TM refer to this period.

Finally, we show how the development for a muon collider is highly synergistic with the developments necessary to other energy frontier collider options, as well as other fields of science and societal applications of high field HTS magnets.

*The magnet challenges for a Muon Collider*

A good impression of the challenges at hand can be gained from the summary of magnet development targets reported in Tab. I. The values there are based on the initial review of the configuration of US-MAP [4], and have evolved in the past years to take into account both the changes in beam physics specifications, as well as the progress in conceptual and engineering magnet design [6, 7]. Tab. I is only a simplified view; it lacks details of the actual machine configuration and aspects such as field quality, which although important, especially for collider magnets, is still under discussion. It is nonetheless useful to give a compact view of the exceptional envelope of performance to be achieved.

At first sight it is already clear that key research and development (R&D) objectives span a very broad range: from achieving very high magnetic field strengths, up to 40 T in solenoids and 14 T in dipoles, to managing stored energies exceeding 1.4 GJ in a single magnet system, mitigating heat loads from nuclear interaction of muon decay at levels of several W/m, and ensuring radiation resistance up to 80 MGy dose. Overcoming these extraordinary challenges requires a paradigm transformation from liquid helium cooled low-temperature superconductor (LTS) magnet technologies operating below 5 K to HTS technologies optimized for efficient operation at temperatures up to 20 K, combined in a compact design to reduce capital expenditure. The concept of a proton-driven muon collider is such that all superconducting magnets listed above operate in steady state, which simplifies operation and opens the possibility of adopting magnet technology such as non-insulated windings that cannot be directly applied to synchrotrons.

Table I. Summary of magnet development targets for the Muon Collider.

| | | Target, decay and capture | 6D cooling | Final cooling | Rapid cycling synchrotron | | Collider ring | | | |
|---|---|---|---|---|---|---|---|---|---|---|
| Magnet type | (-) | Solenoid | Solenoid | Solenoid | NC Dipole | SC Dipole | Dipole | Dipole | Dipole | Quadrupole |
| SC material options | (-) | HTS | HTS/LTS[(2)] | HTS | N/A | HTS | Nb-Ti | $Nb_3Sn$ | HTS | HTS |
| Aperture | (mm) | 1400 | 60...800[(3)] | 50 | 30x100 | 30x100 | 160 | 160 | 140 | 140 |
| Length | (m) | 19 | 0.08...0.3[(3)] | 0.5...1[(4)] | 5 | 2 | 4...6[(4)] | 4...6[(4)] | 4...6[(4)] | 3...9[(4)] |
| Number of magnets | (-) | 20 | 2 x 3030 | 20 | 7000[(6)] | 3000[(6)] | 1250[(8)] | 1250[(8)] | 1250[(8)] | 28 |
| Bore Field/Gradient | (T)/(T/m) | 20 | 2.6...17.9[(3)] | > 40 | ± 1.8[(5)] | 10 | 5 | 11 | 14 | 300 |
| Ramp-rate | (T/s) | SS | SS | SS | 3320...810[(7)] | SS | SS | SS | SS | SS |
| Stored energy | (MJ) | 1400 | 5...75 | 4 | 0.03 | 3.4 | 5 | 20 | 24 | 60 |
| Heat load | (W/m) | 2[(1)] | TBD | TBD | 1200 | 5 | 5 | 5 | 10 | 10 |
| Radiation dose | (MGy) | 80 | TBD | TBD | TBD | TBD | 30 | 30 | 30 | 30 |
| Operating temperature | (K) | 20 | 20 | 4.5 | 300 | 20 | 4.5 | 4.5 | 20 | 4.5...20 |

NOTES:
(1) Intended as linear heat load along the conductor wound in the solenoid. Total heat load in the target, decay and capture solenoid is approximately 4 kW.
(2) Superconducting material and operating temperature to be selected as a function of the system cost. Present baseline study is oriented towards HTS at 20 K.
(3) The range indicated covers the several solenoid magnet types that are required for the cooling cells. Extreme values typically do not occur at the same time.
(4) Specific optics are being studied, the length range indicated is representative.
(5) Rapid Cycled Synchrotrons require uni-polar swing, from zero to peak field. Hybrid Cycled Synchrotrons require bi-polar swing, from negative to positive peak field
(6) Considering the CERN implementation (SPS+LHC tunnels)
(7) Required ramp-rate decreases from the first to the last synchrotron in the acceleration chain
(8) Considering a collider of the final size (approximately 10 km length)

We can elaborate further on the above development targets, and associated challenges:

- The solenoids of the target, decay and capture channel produce a large field, 20 T, in a large bore, 1.4 m, which results in large stored energy and forces. They are subjected to high radiation heat load, of the order of several kW, which calls for operation at temperature ~ 20 K, well above the boiling point of liquid helium, for reasons of thermodynamic efficiency. The main challenges identified for these magnets revolve around:
    - *High current conductor*, 60 kA and higher, suitable for the projected level of forces (mechanics) and stored energy (protection). This class of conductors is actively being developed for fusion applications, but so far, few solutions have proven, at least conceptually, to have the level of robustness required for continuous operation over several years;
    - *Winding technology*. Although the baseline solution is rather standard, and was used successfully in fusion with LTS internally-cooled cables, it was never applied to HTS conductors of this class. Particular attention will be

required to avoid mechanical degradation during winding, as well as finding engineering solutions for the high-current joints and terminations;
- *Radiation resistance*, especially bearing in mind that the physics of radiation damage in HTS is far from being fully characterized and understood. The type of radiation and energy spectrum are very far from those produced by existing and planned test facilities [24]. A sound understanding of the physics of damage will be a mandatory condition to extrapolate from measurements to operating conditions.

- <u>The solenoids in the cells of the 6D cooling channel</u> comprise a very large number of variants, a total just short of 3000 magnets, ranging over a very large span of fields and bore sizes, from combinations of modest value of on-axis field, 2.6 T, and large bore, about 1.5 m, to relatively high values of on-axis field, 17.9 T, and modest bore dimensions, 90 mm. In fact, the on-axis field, size, stored energy and forces vary differently from unit to unit, and it is not possible to select a unique magnet development target for this complex that can be relevant for the whole collider. Furthermore, depending on the geometry and field gradients needed in a cell, the peak field in the solenoid can be much larger than the required on-axis field, adding yet another parameter to the definition of development targets. This is also the reason why the operating temperature has not been set yet, though the preference would be for temperatures above liquid helium for reasons of engineering simplification and economics. Besides the performance of a single solenoid magnet, which can be challenging but feasible, the two main concerns are:
  - *Effective integration of multiple solenoids in the cell lattice*. Integration has geometry and space constraints from other components such as the absorber and RF cavities, structures that sustain the large electromagnetic forces among the solenoids, and effective thermal management;
  - *Quench management* is a challenge, especially in the lattice configuration, including the balancing of the forces among neighboring cells.
  - *Cost of the large-scale production*. The 6D cooling channel has been evaluated as one of the major cost items for a Muon Collider, over one third of the cost of the complete magnet system, which will require aiming at minimum coil dimensions, high engineering current density and careful optimization.

- <u>The final cooling solenoid</u> is well defined in terms of peak field, with a minimum of 40 T, and bore dimension, 50 mm clear ID. Length and configurations are being studied, but it is already clear that the solenoids for the final cooling cells will need a magnetic length in the range of 0.5 m to 1 m. It is likely that this magnet will require low operating temperature, in the vicinity of liquid helium, to increase the operating margin. The main challenges that we have identified for this set of solenoids are
  - *High forces and stresses*, the design hoop stress is in the range of 600 MPa;
  - *Quench management with high energy density in the winding pack*, up to 300 J/cm$^3$;
  - *Novel magnet technology* for winding, splicing, and structural support to cope with the very challenging performance requirements.

- The resistive dipole magnet systems of the Rapid- and Hybrid Cycling Synchrotrons (RCS and HCS) have the largest share in stored energy and losses. They pose the main challenge for this complex. For the resistive dipoles of all RCS's and HCS's we have defined a common specification of field, 1.8 T, and a rectangular aperture, 30 mm(V) x 100 mm(H), compatible with classical design solutions. Still, we require careful optimization of geometry to minimize the stored energy, and evaluation of losses originating from resistive, magnetization, and eddy current effects. In fact, it is the whole system composed of magnet, power converter, and energy storage that needs to be optimized and, eventually, demonstrated for performance and energy efficiency. The challenges identified for this system are:
    - *Efficient storage of the magnetic energy* in the range of several tens of MJ;
    - *Management of reactive power* in the range of tens of GW;
    - *Resistive and magnetic losses* associated with high ramp-rate;
    - *Field quality*, including the effect of all magnetic and conducting materials, and in particular the beam pipe.
- The superconducting dipole magnets of the HCS's has been specified to align with the resistive magnets in terms of aperture, 30 mm(V) x 100 mm(H), while the bore field, 10 T, was selected to keep a sufficiently large palette of design options open. After initial conceptual design considerations, it was realized that these magnets, like the collider magnets (see next points) are subjected to significant radiation heat load and potential radiation damage. We have hence selected an operating point at 20 K to ensure cryogenic efficiency, which implies that magnets must be built with HTS. We recall that these superconducting magnets operate in steady state, which is a significant simplification in terms of powering (no need for ramping), field quality (the field can be stabilized and, possibly, "shimmed") and cooling (no AC loss during operation). A further simplification may come from the fact that the beam residence in the accelerator is very short, typically a few tens of orbits, and field quality may be less of an issue. Although the performance specification and dimensions of these magnets are modest, the main challenge remains:
    - *Reaching the desired field performance with HTS magnet technology*. This challenge is shared with the collider magnets, see below.
- Under collider magnets are several dipole and quadrupole types required for both the arc and the Interaction Regions (IR). A notable requirement for the collider is the need of combined function magnets, dipole and quadrupole as well as dipole and sextupole, to reduce the straight sections to a minimum length. While a complete beam optics solution is still being defined, we have identified specific design points for dipoles and quadrupoles that are likely to be acceptable from the machine point of view but remain greatly challenging from the magnet point of view. The design points, reported in Tab. I, are representative of the technology required for the collider ring: the 10 TeV collider in a 10 km tunnel would require high dipole fields, 14 T, and gradients, 300 T/m, combined with large apertures, 140 mm, to accommodate a radiation shield [25, 26]. As shown in [18], these performance targets trade challenges of comparable difficulty by approaching the engineering boundaries of stress and quench protection. The magnets would be operated above liquid helium conditions to achieve high cryogenic efficiency, ideally in the range of 20 K, unless lower temperatures are required for performance reasons. Like the other sub-systems described earlier, this implies that HTS must be used as the baseline material. Staging options based on LTS are considered, also reported in Tab. I, namely either 11 T Nb$_3$Sn magnets, or 5 T Nb-Ti magnets, both requiring 160 mm aperture to accommodate a larger radiation shield than HTS. These LTS options could be adopted for a 3 TeV collider stage in the final 10 km tunnel (Nb-Ti, 5T), or a 3 TeV stage in a 5 km tunnel (Nb$_3$Sn, 11T), or an approximate 6 TeV stage in the final 10 km tunnel. Like the superconducting dipoles of the HCS, the collider magnets are steady-state, which reduces demands with respect to the main magnets for synchrotrons. The main challenges of collider magnets, dipoles, quadrupoles and combined functions, remain nonetheless many:
- *Reaching the desired field performance with HTS magnet technology*. No accelerator magnet exists at the level of 14 T bore field and large aperture, in the range of 140 to 160 mm.
- *Development of winding technology* for compact, high engineering current density, likely non-planar coils that produce the required field shapes, including ends, joints and terminations;
- *Managing large forces and stresses*, ideally reaching design transverse compression values in the range of the known limit of REBCO, 400 MPa, to keep the coil as small as possible;
- *Quench protection* of magnets with large stored energy, several MJ per unit, and energy density, in the range of 260 J/cm$^3$;
- *Cost*, to achieve convenient economics in the large series of magnets required in the collider. Similar to the solenoids of the 6D cooling channel, the collider magnets for the 10 TeV option, based on HTS, represent about one quarter to one third of the cost of the complete magnet system of a Muon Collider.

While the main focus of this proposal is on the beam guiding magnets of the muon collider, from muon production, through cooling and acceleration, to collision, it is clear that addressing the engineering issues of magnets with large bore and stored energy will also benefit a new generation of detector magnets operated at temperatures above liquid helium (see as an example [27] and [28]).

**State of the art in LTS and HTS magnets of relevance to the Muon Collider**

A complete overview of the state of the art in high field magnet technology, solenoids, accelerator dipoles and quadrupoles, and associated powering systems is out of the scope of this proposal. Here we restrict ourselves to the main achievements in all-superconducting magnets relevant to accelerators, including advances in solenoids of type and dimensions suitable for muon production, capture and focusing. We neglect hybrid superconducting/resistive magnets that are the state-of-the-art in steady state ultra-high fields up to 45 T, because these magnets tend to be large and power hungry, so are not suitable for application in accelerators. However, we will include performance demonstrations consisting of small superconducting coils in a background field, whether the background is provided by superconducting or resistive magnets. We will not enter in details, and rather refer for further information to the extensive literature available. For completeness, we will add considerations on pulsed magnet systems, relevant to the accelerators.

We use here the Technology Readiness Level (TRL) [29] as a global indication of maturity for the deployment in an accelerator project, and quantify the technology gap in terms of the difference to the required TRL. It is commonly accepted that a decision on technology insertion and technology transition requires a TRL 6, i.e. engineering-scale models or prototypes tested in a relevant environment. This is also the level required by the US-DOE to start construction of a new infrastructure [30]. We require a TRL of at least 6 to decide on construction (technology demonstrated in relevant environment). A TRL 6 is the level to be set as the goal for the R&D program. Before industrial procurement a value of at least 7 would be necessary (system prototype demonstration in operational environment). Given the research environment of a particle accelerator, and the moderate competitive industrial environment, a TRL of 8 is appropriate for production (system complete and qualified).

*Solenoids*

The field produced by superconducting solenoids has grown by nearly an order of magnitude from the first attempts in the range of 5 T in experimental solenoids built with Nb-Zr and $Nb_3Sn$ in the early 1960's, to present state-of-the-art just short of 50 T in small demonstrators wound with REBCO. This has been greatly motivated by research needs in fields of material and life sciences.

At the high end of the field reach are the solenoids built with an LTS background coil and an HTS insert. The highest field reached is the 32T all-superconducting user facility at NHMFL (Tallahassee, FL, USA) [31]. The 32T is built as a hybrid with Nb-Ti and $Nb_3Sn$ outsert and HTS (REBCO) insert, it has a clear bore of 32 mm and a stored energy at full field of 8.4 MJ. The magnet was highly successful in reaching full field for the first time in 2017 and operated as a user facility starting in 2019. After initial operations, indications of performance degradation involving the current leads to the quench heaters for the HTS magnet emerged in 2023. Following the repair, the system had a commissioning incident that resulted in damage to the HTS insert coil. The 32T insert is presently under repair with intention to return to operation as a user facility as soon as possible. Although only scantily documented, Chinese researchers at the IPP-CAS (Hefei, China) have achieved similar results, 32.35 T, also using a LTS/HTS solenoid [32]. This class of solenoids are developed and built as one-off realizations, for laboratory applications, and have an estimated TRL of 6 to 7.

The highest field in a commercial solenoid is the one offered by the NMR Ascend 1.2 GHz system of Bruker Biospin [33]. This is a 28.2 T solenoid with a 54 mm bore built with a LTS outsert (Nb-Ti and $Nb_3Sn$) and HTS insert (REBCO). It is to be noted that NMR magnets have exceptional field homogeneity, ppb, and stability, ppb/h, that are much beyond accelerator requirements. Such systems are built as commercial instruments, but in few units. Given the small-scale production, we can estimate a TRL of 8 for this class of solenoids.

On the research side, the highest field reached in an all-HTS solenoid is 26.4 T in a 35 mm bore built as a collaboration of SUNAM, MIT and FSU [34]. The HTS coil uses a "no insulation" (NI) approach where turns are in electrical contact but current does not cross non-superconducting components when in normal operation. This results in an extremely compact magnet running at an engineering current density of 400 $A/mm^2$. The magnet has been tested successfully at SUNAM, and we can assign to this medium-size realization a TRL of 4 to 5.

Even higher field, 45.5 T, was achieved by the Little Big Coil (LBC) solenoids, a series of NI demonstrators with small bore, 14 mm, and outer diameter of 34 mm [35]. A LBC was built assembling double pancakes wound with 4 mm HTS (REBCO) tape. Testing of LBC coils took place in a 31 T background field produced by a resistive solenoid. The record field was reached at a spectacular value of engineering current density of 1420 $A/mm^2$. Although highly successful in demonstrating exceptional potential in the field reach, LBC coils also identified mechanical limitations associated with the effect of shielding currents, combined with specific features of the REBCO tape used to build them. These are demonstration experiments, still rather far from engineered realizations, which is why we assign a TRL of 3 to 4.

It is important to recall here that the results reported above are only a small fraction of the on-going activities using NI winding and its variants (partial-, controlled-, metal-insulated, and other). Indeed, significant work is on-going on solenoid magnets for laboratory application across several groups in academia and industry. Relevant to our discussion, and to cite only a few, are the joint activities at CEA and LNCMI towards 32 T and 40 T hybrid LTS/HTS magnets [36], research at the Applied Superconductivity Laboratory (ASL) at Seoul University, highly active in NI magnet technology [37], and activities at PSI to build a NI solenoid for a positron source [38]. Industry is also highly active, as proven by recent results obtained by Commonwealth Fusion Systems in the US [39], and Tokamak Energy in the UK [40].

The very frontier of DC, ultra-high fields for laboratory applications is embodied by the research work towards 60 T solenoids, first formally advocated by [41] and recently by [42]. These magnets are presently designed as hybrid technology, with a large bore superconducting outsert that provides a background field, and a resistive insert that boosts the field in the small-bore sample space [43]. The highest field achieved with such configuration is 45 T, albeit with large electrical consumption, 20 to 30 MW per single unit, which does not make them well adapted to accelerator applications. The concept, however, could be adapted in the future to make it fully superconducting.

At the high-field end of solenoids with large bore, the magnet with largest stored energy is the ITER Central Solenoid (CS), which produces a central field of 13 T and has a free bore of 2 m [44]. The full CS system consisting of six identical modules, wound with a $Nb_3Sn$ conductor, will have a stored energy of 6.4 GJ. The full CS is not yet assembled, but each single module was tested to the full current 40kA, reaching nominal performance. Indeed, this is confirmation of the finding of the CS Model Coil results, a reduced size solenoid with 1.6 m bore, achieving 13 T, and a stored energy of 640 MJ [45]. Solenoids of this class have been industrially produced, in few units and using custom tooling. We can hence assign them a TRL of 8.

On the research side, activities are on-going on large bore HTS solenoids for fusion. The highest field reached to date in magnets of this class is 5.7 T in a pulsed solenoid with free bore of 1.2 m, storing a magnetic energy of 3.7 MJ [46]. This solenoid was built using an internally cooled and reinforced HTS (REBCO) conductor, and is a steppingstone towards higher fields, 20 T, required by SPARC. We can assign to this laboratory-scale realization a TRL of 5 to 6.

*Accelerator magnets*

As for solenoids, the field produced by dipole and quadrupole accelerator magnets has progressed greatly in the past years, though the absolute increase has been more modest, from the 4.2 T of the Tevatron to the 12 T of HL-LHC. From the first large scale realizations based on Nb-Ti, the Tevatron [47] and HERA [48], accelerator magnets have reached maturity with the LHC [49], which has a nominal field of 8.33 T, two bores of 56 mm diameter, 14.3 m magnetic length and a stored energy of 7 MJ at nominal field. This is the highest performance to be expected from Nb-Ti. Accelerator magnets, dipoles and quadrupoles, built with Nb-Ti, are industrially mature for a large size scientific experiment, i.e. a TRL of 8.

R&D on accelerator magnets built with $Nb_3Sn$ has been ongoing since the mid 1990's, and has led in the past few years to the first realizations that are planned for installation in a running accelerator. They are the 11 T [50] and QXF [51] quadrupoles for the interaction regions of the High-Luminosity upgrade of the LHC (HL-LHC) [52]. These magnets have reached bore fields in excess of 11 T, and peak fields above 12 T. In particular, the HL-LHC 11 T dipoles have a bore of 60 mm, magnetic length of 5.5 m, and a stored energy of 5.5 MJ. The scale of their realization is a few units, with methods that are still appropriate for laboratory scale construction but not for extrapolation to large scale. This is why we assign a TRL of 6 to 7 to this technology.

The highest field achieved with $Nb_3Sn$ conductor in a free bore is that of the FRESCA2 dipole [53], built in collaboration between CERN and CEA and generating a field of 14.6 T in an aperture of 100 mm diameter [61], 1.5 m long and a stored energy of 4.5 MJ. This is not an accelerator dipole and was fully built in a laboratory. The corresponding TRL is hence around 5 to 6.

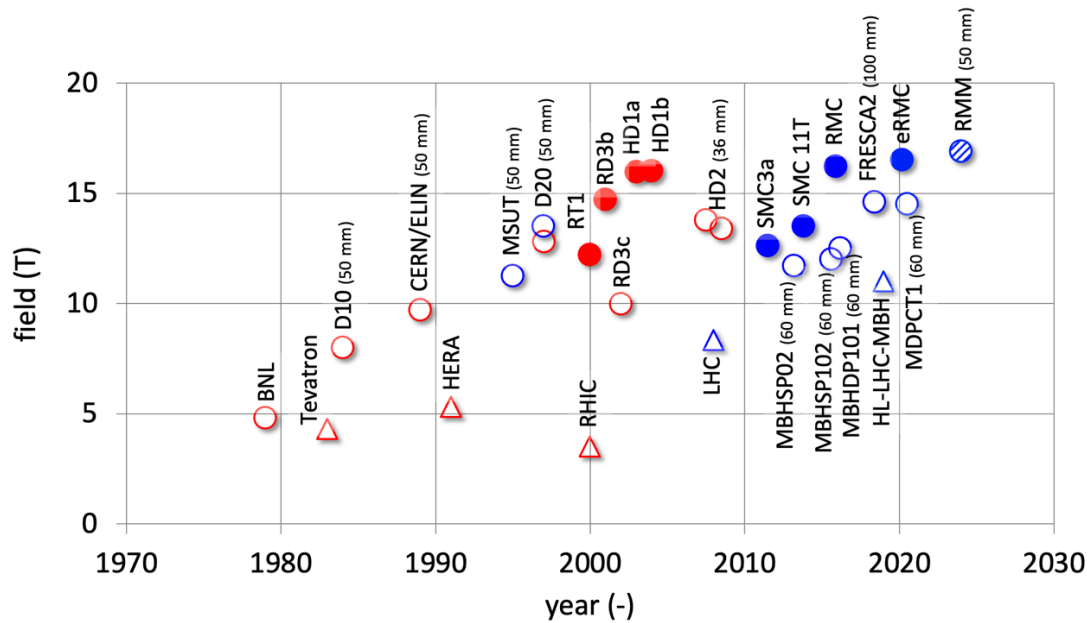

Figure 1. Overview of peak fields achieved with $Nb_3Sn$ dipoles. Closed round symbols represent magnets with no clear bore, open round symbols magnets with bore. Triangles are magnets built for accelerators (note that Tevatron, HERA, RHIC and LHC are built with Nb-Ti). Red symbols indicate operation in the vicinity of liquid helium (4.2 K), blue symbols are for operation in superfluid helium conditions (1.9 K).

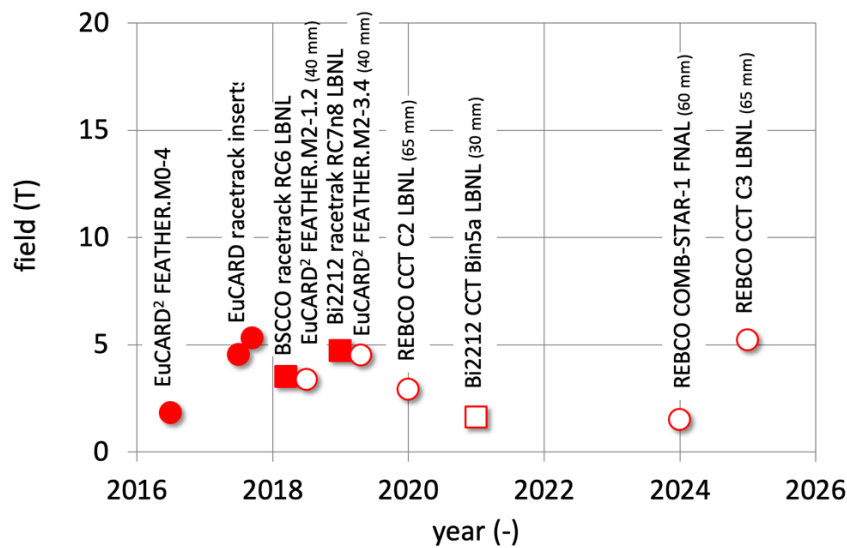

Figure 2. Overview of peak fields achieved with HTS dipoles. Closed round and square symbols represent magnets with no clear bore, open round and square symbols magnets with bore. Round symbols correspond to magnets built with REBCO, square symbols to magnets built with BSCCO-2212. Operation in the vicinity of liquid helium (4.2 K).

To complete the picture of $Nb_3Sn$, R&D coils have reached even higher fields, exceeding 16 T in the Racetrack Magnet Model at CERN [62], recently attaining record fields of 16.9 T. This magnet does not have an aperture, and a total length of about 700 mm. The interest of this test is to probe performance limits of $Nb_3Sn$, which are indeed in the range of 16 T. The TRL is however low, of the order of 4 to 5, since much of the magnet engineering that would be required for an accelerator application is not reflected in these demonstrators. An overview of the maximum fields achieved with $Nb_3Sn$ dipoles is shown in Fig. 1.

As to HTS, the interest in the potential for accelerators began to solidify only in the mid-2000s both in the EU and in the US. Initial efforts in the EU were driven by the interest of a post-LHC collider, within the framework of EU-FP7 EuCARD [53], EuCARD2 [54], EU-H2020 ARIES [55], and US-MDP in the US [56]. Magnets built with REBCO manufactured in Europe and the US, racetracks and block-coil magnets with 40 mm to 65 mm bore and short length, have achieved fields ranging from 3 T to 5 T at 4.2 K and operated stably in gaseous helium up to 50 K and higher at correspondingly reduced field. It is to be noted that magnets with good initial performance often showed degradation attributed to differential thermal and electromagnetic stresses. BSCCO demonstrators of comparable dimensions also achieved comparable fields, up to 4.7 T in the US [57]. HTS accelerator magnets are still at the beginning of their development, not much beyond demonstration, and have an estimated TRL of 3 to 4. As for $Nb_3Sn$, we provide an overview of the maximum fields achieved with HTS dipoles in Fig. 2.

*Pulsed powering system and resistive magnet systems*

The proposed high dB/dt (3300 T/s) resistive magnets with 1.8 T peak field and 11 kA peak current for the muon accelerator share characteristics with existing accelerator technologies but push performance beyond current implementations in a number of aspects.

From the point of view of magnet technology, such magnet performance seems within the reach of existing technology, suitably adapted. Indeed, resistive magnets for accelerators are industry standard, which is why we assign a TRL in the range of 6 to 8. This is a relatively wide range that depends on material selection (e.g. high saturation poles), design features (e.g. control of eddy currents and losses, end effects, shimming for field quality) and specific engineering solutions adopted (e.g. single turn coil manufacturing). The key technical challenges, eddy current effects on the vacuum chamber, and mechanical/thermal stresses due to rapid cycling, are well understood from existing accelerator systems, though they have not yet been demonstrated by a system at this exact field/ramp combination.

Similarly, the power conversion and energy storage principles depend on established concepts which are close to industry standard. With the concept adopted, components are available, and we can assign a similar TRL of 6 to 8 to the powering units, with questions on long term reliability and production scale that justify the low-end of the estimate range.

The challenge is truly the integrated system. The magnet and powering system proposed for the muon collider has similarities with the J-PARC RCS dipoles [58], which operate at 1.1 T peak field, 6.1 kA peak current, and ramp in approximately 20 ms (about 70 T/s) at 25 Hz. The system proposed for the muon collider demands a significantly higher dB/dt (3300 T/s), although at a lower repetition rate (5 Hz) and requires a pulsed, low-duty-cycle power supply, differing from J-PARC's continuous resonant excitation. On the other hand, the system proposed for the muon collider shares features with fast kicker magnets, which achieve even higher dB/dt (>10,000 T/s) but with lower peak fields (0.1 T to 0.5 T). These systems, such as the LHC injection kickers [59], handle peak currents of 3 kA to 10 kA in microsecond-scale pulses, using ceramic vacuum chambers with conductive coatings or strips to mitigate eddy current effects. A similar solution will be necessary for the proposed magnet due to the high dB/dt inside the gap. Based on this assessment, the technology readiness level (TRL) for the system of magnets and powering is an estimated TRL of 4 to 5.

**Gap analysis**

Contrasting the state of the art to the demands and challenges listed earlier, we see that the development of magnets for a muon collider faces several technical and operational challenges, creating significant gaps that must be addressed by focused R&D.

All-HTS solenoids with large bore and high field, such as the concept selected for the target, decay and capture channel, are close to a suitable TRL. This profits from the development of LTS solenoids with large bore and high field, mainly for fusion, which has reached an appropriate TRL of 8 (e.g. ITER magnets). The HTS concept selected uses as much as possible the technical solutions of the LTS counterpart but still needs to progress up from TRL 5 to 6 by at least 1 TRL step, to a secured TRL 6.

In the case of the all-HTS UHF solenoids with NI winding technology, which is promising for the final cooling, the required advances are more substantial. In this case R&D and demonstration is needed to advance from the present TRL values of 3 to 4 to a minimum of TRL 6, by 2 to 3 TRL steps.

The solenoids of the 6D cooling are in the intermediate range of field and bore dimension, with stored energy in the range of several MJ. In this case there is no obvious comparison to the state-of-the-art since no large-size all-HTS magnet has been built. The winding technology being considered for these solenoids, partially insulated, is comparable to that of the UHF experiments, which are at a TRL of 3 to 4. In addition, structural and quench protection issues are made more severe because of the large dimensions, electromagnetic forces, and stored energy. We hence consider that also in this case a progression to a secured TRL of 6 will require an increase of TRL by 2 to 3 steps.

Turning now to accelerator magnets, in the case of Nb-Ti dipoles and quadrupoles, with a TRL of 7 are essentially ripe for industrial production, so no gap can be identified motivating a dedicated R&D or demonstration. For $Nb_3Sn$, the required performance aligns rather well with the achievements of HL-LHC, where the estimated TRL is already close to the desired value of 6. Furthermore, the High Field Magnet R&D programme [62] will continue increasing the TRL of accelerator magnets beyond the HL-LHC benchmark, higher field and forces, thus also preparing solutions suitable for a muon collider. We can hence safely state that also in the case of $Nb_3Sn$ we do not identify a significant TRL gap from the point of view of performance that would motivate large R&D investments. At the same time, a large-scale production of $Nb_3Sn$ accelerator magnets will require full scale prototyping, effective industrialization and cost optimization.

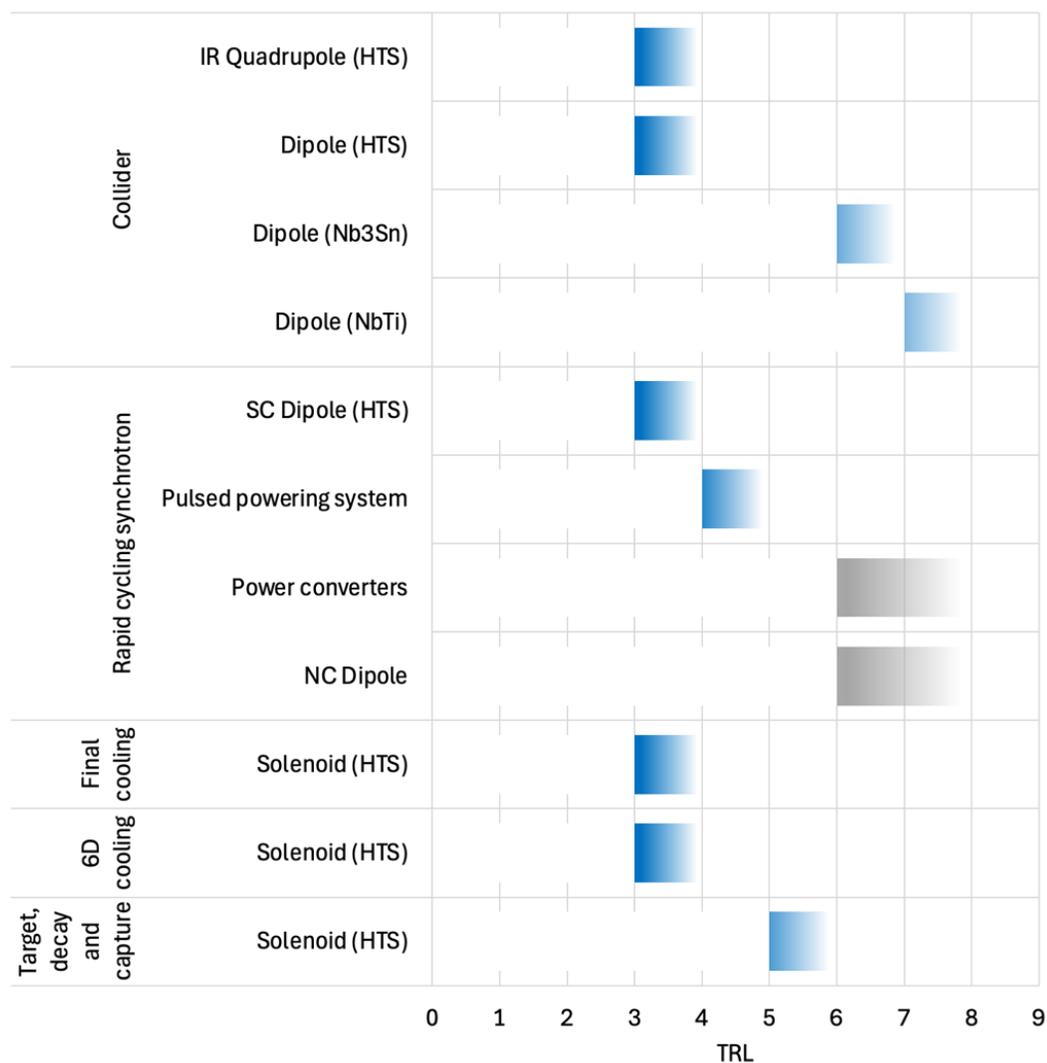

Figure 3. Summary of estimated TRL for the magnets of the muon collider. Normal Conducting (NC) dipoles and power converters for the rapid cycling synchrotrons are quoted separately (grey area) and as a system. A secured TRL of 6 is needed for a decision of construction.

In the case of the HTS accelerator magnets, on the other hand, we are only at the beginning of development and many issues need to be addressed, resolved and demonstrated. A significant increase in TRL is required, by 3 units to reach a secured TRL of 6.

As to the RCS resistive magnets and power converters, component R&D shall focus on prototyping single units (full scaled prototypes for both the magnets and power converter cell with adapted power electronics) to understand and improve the effective operational capability and repeatability of the single units. This will secure a TRL level of 6 for all unit components. In addition, the prototypes will be used to elaborate appropriate reduced scale models to eventually (with a combination of real and simulated systems) build a complete accelerator unit with an appropriate scale factor. Using the scaled unit, a system test (string test) will allow studying the aspect related to control and operation. Besides field reach, energy storage and power efficiency, a system test will provide precious indication on repeatability and tracking. Due to very fast ramping of the field and current, we believe we could use an iterative learning control approach that would rather adapt the current reference from pulse to pulse instead of trying to correct it during the ramp. The concept is very promising particularly given the shape of the pulses is foreseen to be the same from pulse to pulse. Such tests are required to advance the TRL of the magnet and powering system by 1 to 2 steps to the required level of TRL 6.

In summary, we report in Fig. 3 a graphical representation of the present and desired TRL for the various magnet and powering systems of a muon collider. We can use this representation to define the scale and type of R&D required, whether still at the level of small-scale demonstrators and models, when the estimated TRL is at the level of 4 to 5, or rather towards full-scale prototypes, for values of TRL 6 and higher. The proposed R&D activities are described in the next section.

**Technology Driven R&D**

We have built the proposal of the R&D necessary to advance the TRL of the muon collider magnet concepts around eight Technology Milestones (TMs). The activities in the TM's address the challenges identified, and are intended to fill the gap identified earlier in this proposal, from present state-of-the-art to a secured TRL 6. In practice, each TM is associated with a magnet demonstrator, except for the RCS-String which is a system test. Achieving the TM corresponds to the construction and successful test of the associated magnet demonstrator or system. As will be seen below, the demonstrators span a broad range of configurations, operational temperatures, and performance targets, covering all unique aspect of the collider's infrastructure. In addition to the eight TM's, one R&D activity is proposed to cover material testing necessary and relevant to the specific needs of the magnets for the muon collider.

While fully relevant and specifically tailored to the muon collider, the proposed R&D program would unfold within the scope of wider R&D activities with connections and implications to other HEP R&D, as well as R&D in other fields of magnet technology and applied superconductivity. Indeed, the TM's are highly aligned with the perspective provided by the US-MDP plans [56], the EU LDG Accelerator R&D Roadmap [62], the recommendations from the US HEP P5 panel [63], the analysis and recommendations from the US National Academy of Science, Engineering and Medicine [42], and the US-DOE Office of Fusion Energy Sciences vision [64]. This is why we have indicated in the detailed description potential synergies and collaborations, as well as the opportunities for co-funding from sources other than the Muon Collider activities.

*TM1. 20 T at 20 K target solenoid model coil (20@20)*

<u>Objective</u>. This technology milestone consists of developing conductor, winding and magnet technology suitable for a target solenoid. The main objective is to demonstrate stable operation and quench protection at the design point.

<u>Demonstrator</u>. The TM is associated with a model coil generating a bore field of 20 T, and operating at a temperature of 20 K, hence the acronym "20@20". The geometry of the model coil is a solenoid built in two equal modules, with approximate 1 m bore size, 2.3 m outer diameter and 1.5 m height. These dimensions are scaled from the geometry of the target solenoid, to reduce conductor material needs and cost. We have taken care to maintain crucial performance parameters identical, or at least fully relevant to the final coil design. One of the configurations being studied is shown in Fig. 4.

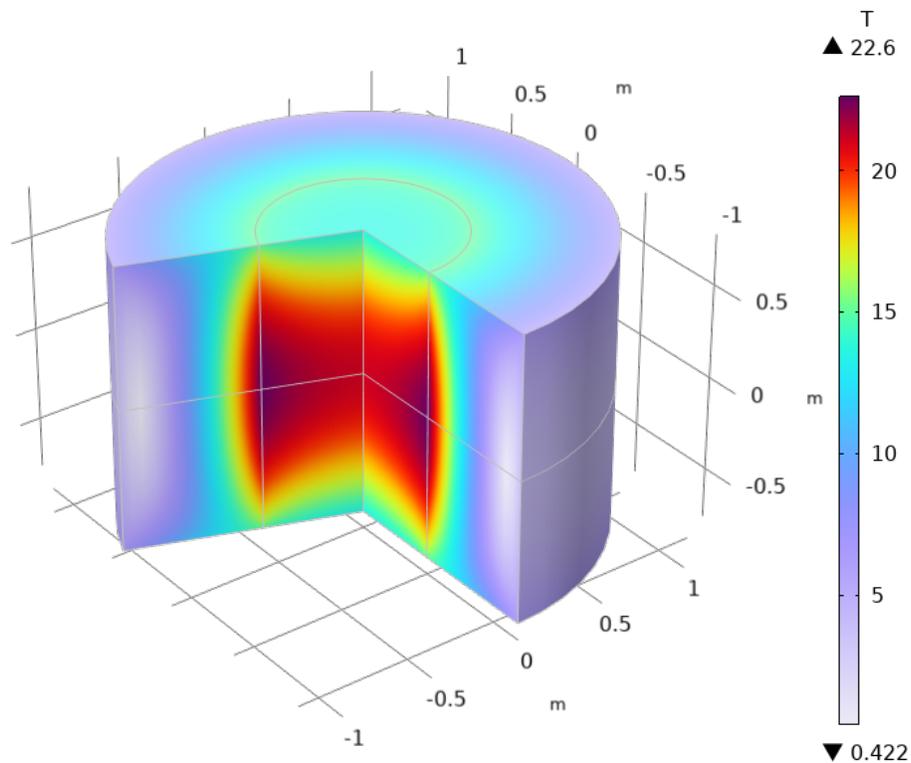

Figure 4. A HTS 20@20 model coil, showing the solenoid and the computed field map.

Timeline and Resources. The estimated time to reach this milestone is 8 years, for a total cost of approximately 30 MCHF, and personnel needs of 37 FTEy. Details on the estimated personnel and material effort are reported in the summary table in Appendix I.

Synergies and collaborations. The model coil described here has attracted significant interest from other fields of applications of HTS magnet technology, e.g. fusion. Indeed, private fusion initiatives are pursuing similar objectives, and the realization of this model coil will likely profit from collaboration and external contributions. With a relatively high TRL for this magnet technology, it is also important that industry has significant participation, also in preparation of future realizations. The work proposed will hence focus on engineering and technical coordination. As a final remark, given the dimensions and stored energy, HEP detectors may also benefit from this technology development.

*TM2. Split solenoid integration demonstrator for the 6D cooling cell (SOLID)*

Objective. This technology milestone consists of demonstrating operating field performance in geometrical and operating conditions relevant to the 6D muon cooling [65]. The main objective is to demonstrate successful integration.

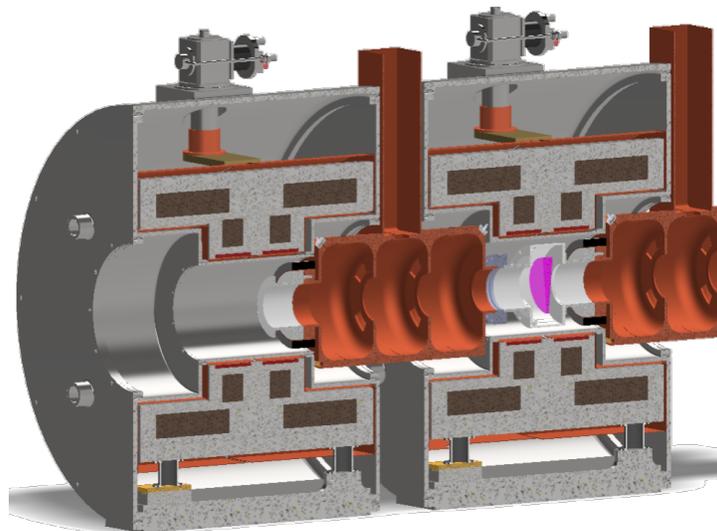

Figure 5. Configuration considered for the study of a Split sOLenoid Integration Demonstration (SOLID).

Demonstrator. The TM is associated with a HTS split solenoid, named SOLID, with target field on axis of 7 T, a free bore of 510 mm, a split gap of 200 mm, operating at 20 K. The performance demands are in the middle of the wide range of solenoid designs required by the 6D cooling channel. The split solenoid demonstrator includes integration in its support structure, being submitted to mechanical and thermal loads representative of a 6D cooling cell. This demonstrator is an intermediate step, bridging the gap between the on-going realization of a test facility for RF cavities in field (RFMFTF), made with up to two HTS solenoids with reduced gap and performance, and the successful production of a full 6D cooling cell, which will require achieving full field performance and respecting nominal space constraints.

The reference configuration is that of the demonstrator cooling cell, shown in Fig. 5.

<u>Timeline and Resources</u>. The estimated time to reach this TM is 7 years, for a total cost of approximately 7 MCHF, and personnel needs of 42 FTEy. Details on the estimated personnel and material effort are reported in the summary table in Appendix I.

<u>Synergies and collaborations</u>. The present TRL for this magnet technology is relatively low, 3 to 4. It is hence likely that most of this manufacturing and testing will take place in research institutes and laboratories.

*TM3.Final cooling UHF solenoid demonstrator (UHF-Demo)*

<u>Objective</u>. This technology milestone consists of demonstrating operating field performance relevant to the final muon cooling solenoid. The main objective is field level attained.

<u>Demonstrator</u>. The TM is associated with a prototype of the HTS final cooling solenoid, named UHF-DEMO, targeting a field of 40 T in a 50 mm bore, and total length of 150 mm. This is shorter than the nominal length of the final cooling solenoid, but given the present modular design this is an inessential point for performance demonstration. This solenoid will likely operate in the vicinity of liquid helium temperature, but will be tested under a wide temperature range to explore performance limits.

The reference configuration is that of the final cooling solenoid, shown in Fig. 6.

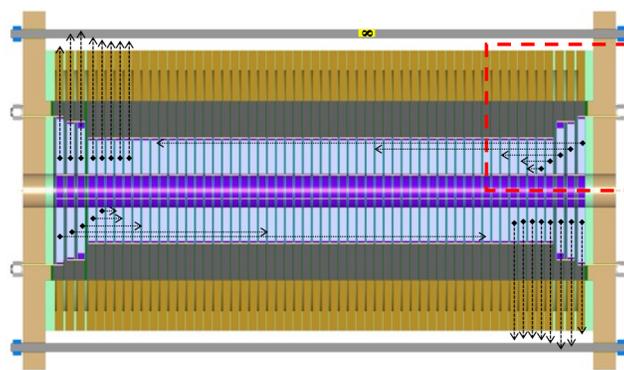

Figure 6. Configuration considered for the study of a UHF solenoid DEMOnstrator (UHF-DEMO).

<u>Timeline and Resources</u>. The estimated time to reach this TM is 9 years, for a total cost of approximately 5.6 MCHF, and personnel needs of 52 FTEy. Details on the estimated personnel and material effort are reported in the summary table in Appendix I.

Synergies and collaborations. The present TRL for this magnet technology is relatively low, 3 to 4. It is hence likely that most of this manufacturing and testing will take place in research institutes and laboratories. This TM pushes the boundaries of current superconducting technology, and is of relevance not only to beam focusing in particle accelerators but also in other fields of frontier science. It is hence likely to attract interest and collaboration from other fields of scientific application like material and life sciences in high magnetic fields and NMR.

*TM4. RCS magnet String and power systems (RCS-String)*

Objective. This technology milestone consists in demonstrating operation of a fast pulsed, energy recovery resistive magnet circuit, including the powering and energy storage infrastructure of the type designed for the rapid cycled synchrotrons. The main objectives are field tracking, field quality and energy efficiency.

Demonstrator. The TM is associated with a string made of resistive dipole magnets, generating a nominal field of 1.8 T in a 30×100 mm aperture. Four such magnets, each 5 m long, are powered in series by a fast pulsed power converter that includes energy storage in capacitor banks that store a total energy of 150 kJ. The target is to reach 3.3 kT/s. Both single-sided and double-sided field swings will be tested, relevant to the RCS and HCS configurations. The aim is to achieve energy recovery efficiency at levels much better than 1 %. The string test will require prior construction and test of the magnets, as stand-alone, and of the powering cells, on a dummy load.

Timeline and resources. The estimate time to reach this TM is 7 years, for a total cost of approximately 6 MCHF, and personnel needs of 20 FTEy. Details on the estimated personnel and material effort are reported in the summary table in Appendix I.

Synergies and collaborations. The present TRL for this magnet and powering technology is relatively high, above 7. Nonetheless, the configuration is very specific to the accelerator application, and it is hence likely that it will be fully realized in research institutes and laboratories.

*TM5. Wide-aperture, steady state $Nb_3Sn$ dipole (MBHY)*

Objective. This technology milestone consists in demonstrating dipole performance at the level required for the muon collider ring, based on LTS ($Nb_3Sn$) technology. The main objective is to demonstrate the combination of field, aperture, training memory and field quality in conditions relevant to the collider operation.

Demonstrator. The TM is associated with a full-size prototype of the $Nb_3Sn$ collider dipole, named MBHY, targeting a field of 11 T in a 160 mm bore, and total length of 5 m. The operating point is in the range of 4.5 K.

Timeline and resources. The estimatd time to reach this TM is 11 years, marginally beyond the horizon of this proposal, for a total cost of approximately 11 MCHF, and personnel needs of 71 FTEy. This includes the necessary small demonstrator coils and models towards the full-size dipole. Details on the estimated personnel and material effort are reported in the summary table in Appendix I.

Synergies and collaborations. The TRL for this magnet is relatively high, estimated at above 6 because of the previous efforts taken for the construction of the HL-LHC 11T dipoles and QXF quadrupoles, which has been done in part in collaboration with EU industry. In fact, development of $Nb_3Sn$ dipoles is within the scope of the High Field Magnet R&D programme, with higher field but smaller aperture. Furthermore, US-MDP is developing large aperture $Nb_3Sn$ magnets for hybrid LTS/HTS R&D. These magnets have specifications comparable to the MBHY demonstrator. We hence expect that this TM can be shared, if not fully delegated, to future activities in the scope of HFM and US-MDP.

*TM6. Rectangular aperture HTS dipole (MBHTS)*

Objective. This technology milestone consists in demonstrating dipole performance at the level required for the muon accelerator hybrid cycled synchrotron (HCS) ring, based on HTS (REBCO) technology. The main objective is to demonstrate the combination of field, aperture and field quality in conditions relevant to the accelerator operation.

Demonstrator. The TM is associated with a model of the HTS accelerator dipole, named MBHTS, targeting a field of 10 T in a rectangular aperture of 100 mm width by 30 mm height, and total length of 1 m. The operating point is in the range of 20 K.

The reference configuration is being studied as part of the design of the steady state accelerator magnets for the HCS, shown in Fig. 7.

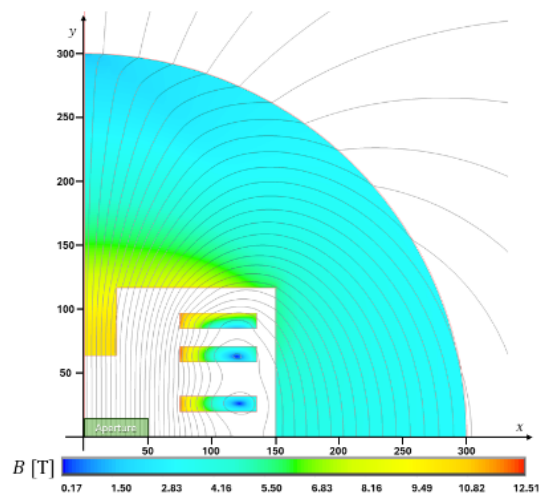

Figure 7. Configuration considered for the study of a 10 T HTS, 30 mm x 100 mm aperture dipole for the hybrid cycled synchrotrons (MBHTS).

Timeline and resources. The plan to reach this TM has been set over a duration of 10 years, for a total cost of approximately 8.3 MCHF, and personnel needs of 60 FTEy. This includes the necessary small demonstrator coils towards the model dipole. In fact, because the demonstrator has a relatively simple geometry and modest dimensions, this development could be completed on a shorter time scale, but results may not be required much earlier than planned. Details on the estimated personnel and material effort are reported in the summary table in Appendix I.

Synergies and collaborations. The TRL for this magnet is low, estimated at 3 to 4. We expect that most of the manufacturing and testing will take place in research institutes and laboratories. This TM has however a wider interest in that it can serve as a testbed for high-field dipole technology while minimizing cryogenic requirements, a major cost driver in collider operations. As such, it could attract interest well beyond the muon collider efforts.

*TM7. Wide aperture HTS dipole (MBHTSY)*

Objective. This technology milestone consists in demonstrating dipole performance at the level required for the muon collider ring, based on HTS (REBCO) technology. The main objective is to demonstrate the field performance, in combination with a wide aperture, including field quality and stability in conditions relevant to the collider operation.

Demonstrator. The TM is associated with a model of the HTS wide-aperture collider dipole, named MBHTSY, targeting a field of 14 T in an aperture of 140 mm, and total length of 1 m. The operating point is in the range of 20 K.

Configurations being studied for the collider dipole are shown in Fig. 8.

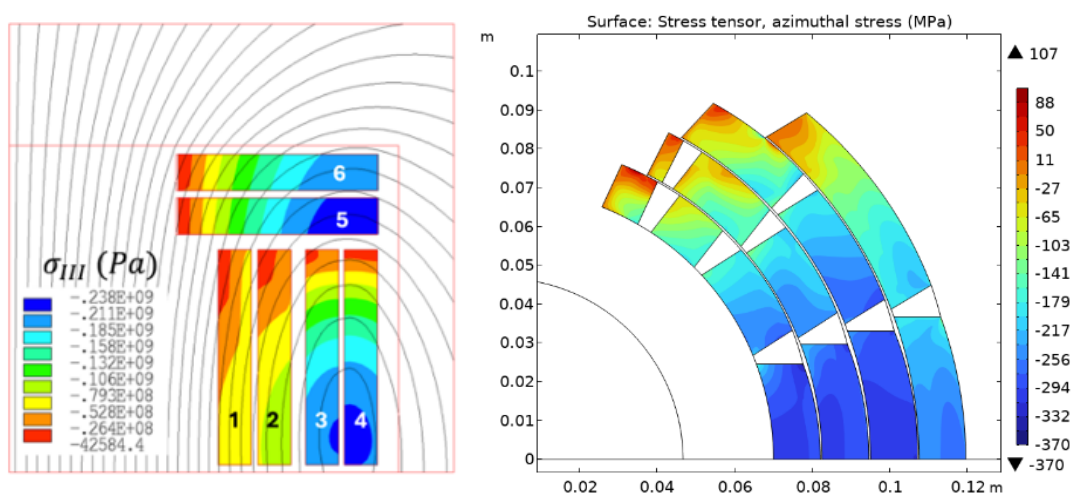

Figure 8. Coil configurations considered for the study of a wide aperture HTS dipole (MBHTSY).

Timeline and resources. The estimated time to reach this TM is 16 years, i.e. beyond the time span considered for the evaluation of resources. The total cost is estimated at 15.8 MCHF, and personnel needs of 126 FTEy. Over the period of interest in this proposal, i.e. ten years, the total cost is estimated at 7.9 MCHF, and personnel needs of 75 FTEy. Details on the estimated personnel and material effort are reported in the summary table in Appendix I.

Synergies and collaborations. The TRL for this magnet is low, estimated at 3 to 4. We expect that most of the manufacturing and testing will take place in research institutes and laboratories. This development targets applications requiring a combination of strong magnetic fields and substantial beam clearance, emphasizing field homogeneity and thermal robustness. Advances in the HTS development produced by the High Field Magnet R&D programme and US-MDP will feed the engineering and construction of this demonstrator. At the same time, we expect that the results achieved will elicit substantial interest from accelerator applications other than the muon collider, such as the FCC-hh.

*TM8. Wide aperture HTS IR quadrupole (MQHTSY)*

Objective. This technology milestone consists of demonstrating quadrupole performance at the level required for the muon collider interaction region, based on HTS (REBCO) technology. The main objective is to demonstrate that the gradient and aperture can be achieved, possibly operating at lower temperature than the arc to increase operating margin, as collider luminosity depends critically on the performance of the IR quadrupoles.

Demonstrator. The TM is associated with a model of a HTS wide-aperture interaction region quadrupole, named MQHTSY, targeting a gradient of 300 T/m in an aperture of 140 mm, and total length of 1 m. Note that, as remarked earlier, these targets correspond to mechanical stresses and quench protection challenges comparable to the large aperture dipole MBHTSY. The operating point is up to 20 K, although lower operating temperature will be explored if necessary to reach the desired performance requirement.

Timeline and resources. The estimated time to reach this TM is 16 years, i.e. beyond the time span considered for the evaluation of resources. In fact, it is planned to start this development only in a second stage, also profiting from the initial results of the large aperture HTS dipole MBHTSY. The total cost is estimated at 8.8 MCHF, and personnel needs of 60 FTEy, assuming that much of the technology will be shared with the development of the wide aperture HTS dipoles MBHTSY. Over the period of interest in this proposal, i.e. ten years, the total cost is estimated at 3.5 MCHF, and personnel needs of 27 FTEy. Details on the estimated personnel and material effort are reported in the summary table in Appendix I.

Synergies and collaborations. The TRL for this magnet is low, estimated at 3. We expect that most of the manufacturing and testing will take place in research institutes and laboratories. As for the large aperture dipole MBHTSY, there are clear synergies with the High Field Magnet R&D programme and US-MDP.

*Materials and methods R&D*

<u>Objective</u>. This line of activity comprises material testing as well as the development of design and manufacturing methods relevant and specific to the magnets for a muon collider. This activity complements other activities of similar nature and scope that take place in companion programmes, such as the High Field Magnet R&D programme and US-MDP. The objective is to support development and testing that are typically shared among several R&D efforts in this proposal, and specific to the developments requested to achieve the TM's in this proposal.

<u>Scope</u>. The material testing and methods development provisionally included in this activity are:

- High-field measurement of transport properties of REBCO conductors, also necessary to define scaling laws required for the design and analysis of the magnet demonstrators;
- Micrography, Micro-structure and mechanical properties of REBCO conductors and winding;
- Radiation effects in REBCO conductors;
- Tailored experiments to establish design rules for HTS magnets, e.g. allowable hot-spot temperature, or allowable peak stress and strain;
- Multi-physics modeling of transient electromagnetics, mechanics and thermal fields in HTS magnets, relevant to the electro-mechanical design, operation and quench protection of NI HTS magnets.

<u>Timeline and resources</u>. The resources estimated for this activity over the reference period of 10 years is a total of 3 MCHF and 30 FTEy. Details on the estimated personnel and material effort are reported in the summary table in Appendix I.

<u>Synergies and collaborations</u>. The testing and development methods which are part of this R&D are relevant and impact other research on HTS magnet technology, for HEP, as well as other applications. It is likely that the results will attract considerable attention and collaboration from within HEP and other fields of scientific and societal applications. This is the area where we expect to work in close collaboration and complementarity with the running High Field Magnet R&D programme and US-MDP. We stress again that this R&D will provide the funding and focus necessary to answer questions specific to the muon collider.

*Summary of Proposed Research and Development*

A summary of the personnel resources and material cost is shown in Figs. 9 and 10, where we report both the yearly values (in FTE and MCHF/y), as well as the cumulated values (in FTE y and MCHF). In the case of personnel, we also report the estimate of the effort that would be required to progress to the next phase of prototyping and construction, should the project proceed in this direction.

The total personnel required for the proposed magnet R&D program, over the ten year period, is 414 FTE y. This is rather evenly distributed, reaching 199 FTE y after five years, and a maximum just above 50 FTE in the middle of the ten year period. To be noted that towards the second half of the ten years of R&D, proficient personnel could transfer to prototyping and construction, seamlessly absorbed in the 60 to 65 FTE that would be needed in total (including residual R&D) during this phase.

Under materials we list all direct raw materials cost, consumables and services performed under contract. We include tooling in the material costs, but we exclude the cost of generic infrastructure (e.g. test infrastructure). The total material costs over ten years are estimated at 82.5 MCHF, reaching 39 MCHF after 5 years, nearly half. The peak yearly expenditure, in the middle of the ten year period, is at the level of 10 to 13 MCHF/year. Note that in this case the drop in yearly expenditure is due to the fact that a transition to the following phase of prototyping and construction is not added.

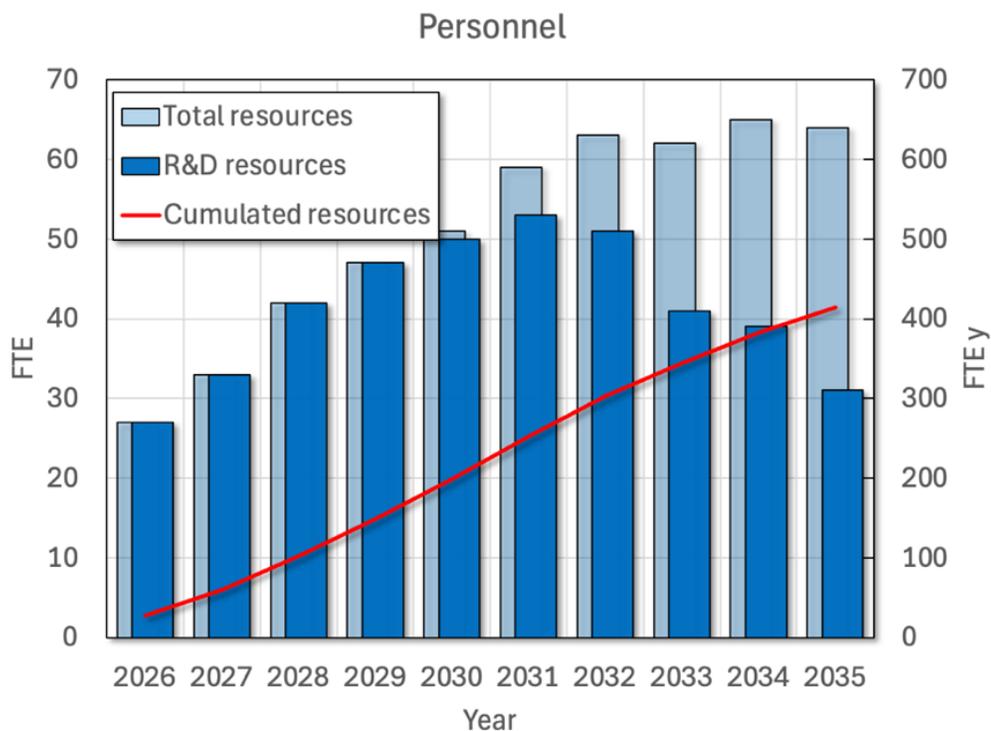

Figure 9. Time profile of the personnel resources necessary to conduct the R&D program proposed in this note, reported both as yearly values (columns, left axis), as well as cumulated over the years of activity (line, right axis). The second set of columns (shaded) reports the estimated total personnel resources that would be needed to proceed to prototyping and construction in the period after 2031.

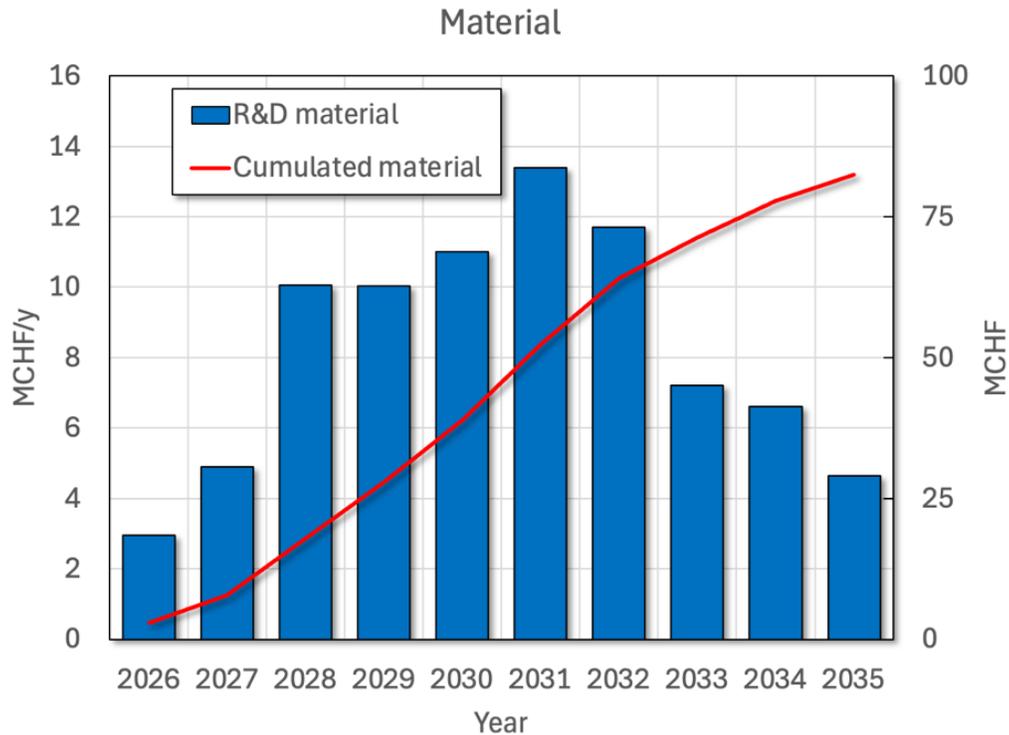

Figure 10. Time profile of the material cost necessary to conduct the R&D program proposed in this note, reported both as yearly values (columns, left axis), as well as cumulated over the years of activity (line, right axis).

**Synergies and collaborations**

We have anticipated that the R&D proposed has the potential for significant impact in fields of science and societal applications well beyond the muon collider. We give in Fig. 11 a synoptic view of our evaluation of the impact. The representation in Fig. 11 is purely graphical, but suggestive, demonstrating that all proposed TM's (rows) have a connection to other programmes in HEP and other fields of application (columns). We detail below the present or anticipated links.

*Other R&D and studies in HEP*

It is evident from the material presented so far that the muon collider magnet development has unique challenges and aspects that go well beyond the targets and R&D of existing programmes, such as the High Field Magnet (HFM) R&D programme, the US-based Magnet Development Programme (US-MDP) and the FCC-hh magnet development. Nonetheless, advancing accelerator magnet technology as proposed here would produce great potential for increased performance or significant construction and operation costs. A relevant example is the FCC-hh. We are aware that the challenges in the synchrotron magnets of a FCC-hh would require development beyond the proposed dipoles and quadrupoles (TM6, TM7 and TM8). Still, demonstrating all-HTS accelerator magnets, built with compact coils at high engineering current density and operated at 20 K would represent a significant paradigm shift.

At the same time, the proposed R&D TM's will profit from advances fostered by HFM and US-MDP in various areas, such as:

- HTS material procurement and characterization;
- cryogenic technology development towards supercritical force-flow cooling, conduction cooling and dry magnets;
- HTS winding technology for solenoids, dipoles and quadrupole magnets;
- analysis and measurement methods;
- test facilities.

In fact, the MBHY demonstrator in TM5, a wide-aperture $Nb_3Sn$ dipole, is a crucial step in the development of LTS/HTS hybrid magnets in HFM and US-MDP. This R&D could be performed jointly, if not fully delegated.

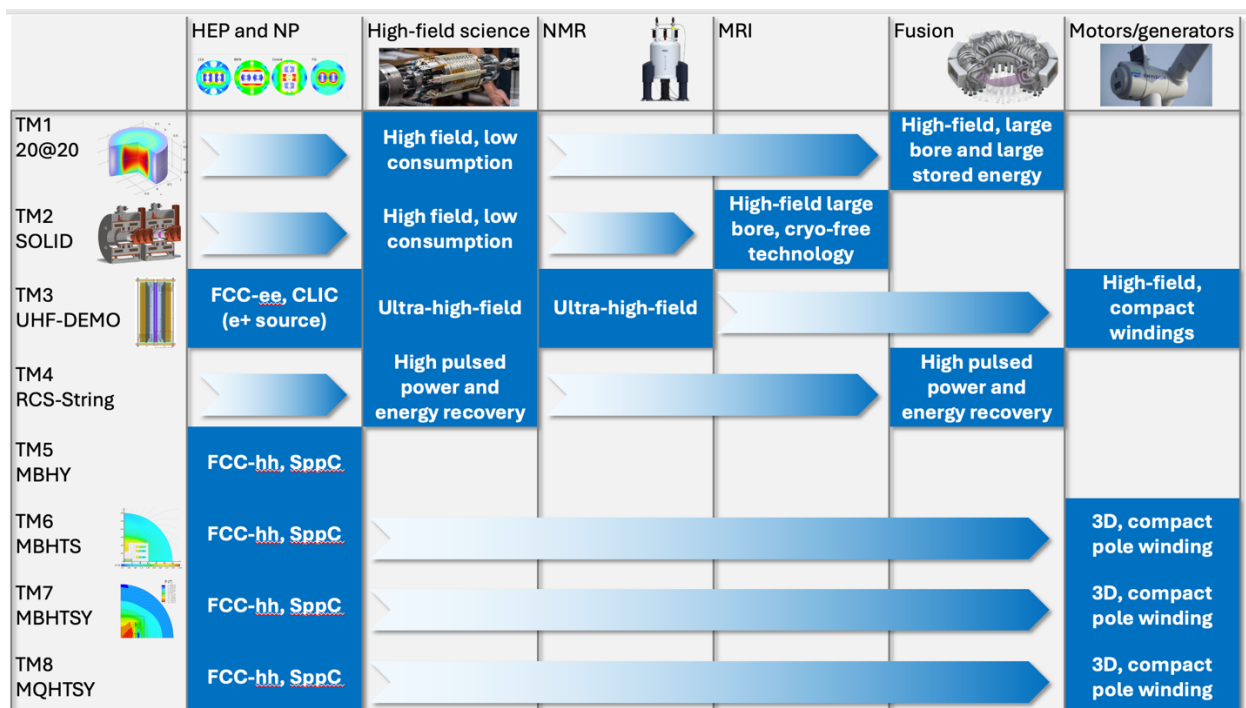

Figure 11. Schematic representation of the impact of the R&D driven by the technology milestones defined in this proposal on other programmes in HEP and other fields of application. The specific technologies of interest are indicated in the highlighted boxes.

*Materials and life sciences*

Advancing the HTS technology for high- and ultra-high-field solenoids (TM1 and TM3) will enhance the capability to study materials under extreme conditions. This is relevant to the creation and characterization of new quantum states, measurement of electronic properties of components such as those in quantum computers, the discovery of new materials, including novel superconductors, or the development of nanostructures as required to increase processing power and storage in next generation electronics, or more efficient batteries.

Higher field will also augment the diagnostic power of NMR in structural and functional inorganic and organic chemistry, and biology. Relevant advances enabled by higher NMR fields include understanding biological processes, mapping protein structures of diseases necessary to develop treatments, or monitoring pollutants in soil and water, including their origin.

The development of solenoids at the upper end of field and aperture at lower electrical consumption than presently possible will facilitate progress in magnet technology that is required for analytical research infrastructure for materials and life sciences, such as the European Magnetic Field Laboratory (EMFL) or the US National High Magnetic Field Laboratory (NHMFL).

High magnetic fields made possible by reaching the goals of this R&D will also improve analysis power and tune-ability of undulators for Light Sources (LS's) and Free Electron Lasers (FEL's) [66] and extend the reach of neutron spectroscopy instruments for the analysis of the structure of matter [67].

*Healthcare*

HTS solenoid technology of the type envisaged in the 6D cooling cells (TM2) will play a major role in the next step in research MRI (Magnetic Resonance Imaging) [68], increasing resolving power, and enabling higher resolution images. HTS magnets with simplified cryogenics, possibly cryo-free, will reduce healthcare expenses and facilitate market penetration of clinical MRI, which still has not reached full exploitation for world regions where high-technology support is not readily available.

Compact, medium field, low consumption HTS accelerator magnets (TM6) would benefit accelerators and gantries for particle therapy, decrease their size and electrical consumption, thus facilitating the diffusion of advanced cancer treatment centers. The therapy accelerators and gantries require rapidly pulsed fields, which is not addressed by the R&D proposed here. Still, steady state magnetic configurations can be envisaged for both beam acceleration (synchrotrons) and gantries, and they would benefit from the advances in the R&D proposed.

*Energy, power and mobility*

Widespread introduction of HTS materials in high field magnets with large bore and stored energy can lead to simpler and more efficient fusion plants. Higher field can be used to massively increase the fusion power, for a given plant size. At the same time, simplified cooling, operating at 20 K, reduces the thermal shield and space requirements, making the overall plant smaller and easier to operate. Indeed, HTS magnet technology may be crucial to resolving the long-standing question on whether thermonuclear fusion will ever be a reliable source of energy for humanity. Inspired by the advances in fusion, the contribution of the R&D proposed here is to demonstrate (TM1) and develop (TM2) HTS magnet technology applicable to magnets with large dimensions and stored energy, i.e. relevant knowledge for fusion applications.

HTS magnet technology with compact 3D polar windings (TM6, TM7) producing high fields (TM3) will benefit generators and motors, increasing their power density, reducing their weight, and making them more efficient. Even if higher energy efficiency is only a marginal benefit, the increased power density and reduced mass allows for more powerful machines which are either not possible or less complex from the engineering point of view. One such example is the generators used in windmills, greatly benefitting from a reduced nacelle mass. A lighter nacelle allows building a higher mast and larger rotor blades, thus also increasing the generator power.

*Other technological and societal impacts*

Among the many technological advances that will be brought by the proposed R&D programme, developing solutions for cryogenic operation at temperatures above liquid helium is one of the most crucial and impactful. Helium cooling of magnets at 20 K, if suitably designed, can improve energy efficiency by a factor of four with respect to operation at 4.2 K, and reduces the complexity and costs of cryogenics systems. This would enable broader scientific, societal and industrial applications of superconducting magnets.

Finally, and more in general, a novel HTS magnet technology can create opportunities for spin-offs, new industrial applications, associated profit and job creation, as well as providing a breadth of opportunities for education and training playground. Promotion of an innovation ecosystem is crucial to drive high-tech economic growth and sustainability initiatives. The significant knock-on effect was quantified in studies of high-tech realizations such as the LHC and the HL-LHC [69].

**Appendix I. Detailed deliverables, personnel and material needs for the proposed R&D program**

Table AI.1. Summary of objectives, main deliverables, personnel and material needs for the 20 T at 20 K (20@20) target solenoid model coil.

| Objectives | | | | | | | | | | |
|---|---|---|---|---|---|---|---|---|---|---|
| Develop conductor, winding and magnet technology suitable for a target solenoid, generating a bore field of 20 T, and operating at a temperature of 20 K. The geometry is based on a model coil, a single solenoid coil with reduced bore size and height, scaled to reduce conductor needs and cost | | | | | | | | | | |
| **High-level Deliverables** | | | | | | | | | | |
| 1) HTS conductor, designed, manufactured and tested on industrial scale for force flow-cooled large bore high field solenoids (1 km) (3Y) <br> 2) Reduced scale windings of final conductor, designed and manufactured with industrial participation, tested in self- and background field (5Y) <br> 3) Model coil, designed and manufactured with industrial participation, tested for performance and endurance (8Y) | | | | | | | | | | |
| **Resources** | 2026 | 2027 | 2028 | 2029 | 2030 | 2031 | 2032 | 2033 | 2034 | 2035 |
| Staff | 0.6 | 0.9 | 0.9 | 1.5 | 3 | 4 | 3.5 | 2.1 | | |
| Postdoc/GRAD | 0.8 | 1.2 | 1.2 | 2 | 1.8 | 2.4 | 2.1 | 0.6 | | |
| Student | 0.6 | 0.9 | 0.9 | 1.5 | 1.2 | 1.6 | 1.4 | 0.3 | | |
| Material | 1000 | 2000 | 5000 | 4000 | 5000 | 7000 | 5000 | 1000 | | |
| **Interested partners** | | | | | | | | | | |
| Academia: CERN, INFN, University of Bologna, Politecnico of Torino, University of Twente, EPFL/SPC, KEK <br> Industry: Tape manufacturers, ASG, ICAS | | | | | | | | | | |

Table AI.2. Summary of objectives, main deliverables, personnel and material needs for the split SOLenoid Integration Demonstrator for 6D cooling cell (SOLID).

| Objectives | | | | | | | | | | |
|---|---|---|---|---|---|---|---|---|---|---|
| Demonstrator of HTS split solenoid performance, including integration in its support structure submitted to mechanical and thermal loads representative of a 6D cooling cell. Target field 7 T, bore 510 mm, gap 200 mm, operating at 20 K | | | | | | | | | | |
| **High-level Deliverables** | | | | | | | | | | |
| 1) Split HTS solenoid design completed (1Y) | | | | | | | | | | |
| 2) Small scale solenoid demonstration tests, validating technology selections (3Y) | | | | | | | | | | |
| 3) Split solenoid built and tested (7Y) | | | | | | | | | | |
| **Resources** | 2026 | 2027 | 2028 | 2029 | 2030 | 2031 | 2032 | 2033 | 2034 | 2035 |
| Staff | 0.9 | 2.1 | 2.4 | 2.4 | 2.1 | 2.5 | 2 | | | |
| Postdoc/GRAD | 1.2 | 2.8 | 3.2 | 3.2 | 2.8 | 1.5 | 1.2 | | | |
| Student | 0.9 | 2.1 | 2.4 | 2.4 | 2.1 | 1 | 0.8 | | | |
| Material | 400 | 900 | 1400 | 1700 | 1200 | 1000 | 500 | | | |
| **Interested partners** | | | | | | | | | | |
| Academia: INFN, CERN, University of Southampton, Technical University Tampere | | | | | | | | | | |
| Industry: Tape manufacturers | | | | | | | | | | |

Table AI.3. Summary of objectives, main deliverables, personnel and material needs for the Final cooling UHF solenoid demonstrator (UHF-Demo).

| Objectives | | | | | | | | | | |
|---|---|---|---|---|---|---|---|---|---|---|
| Build and test a demonstrator HTS final cooling solenoid, producing 40 T in a 50 mm bore, and total length of 150 mm (limit cost) | | | | | | | | | | |
| **High-level Deliverables** | | | | | | | | | | |
| 1) Single pancake, final configuration, stand-alone test (2Y) | | | | | | | | | | |
| 2) Stacked pancake, final configuration, achieve 20 T (5Y) | | | | | | | | | | |
| 3) Demonstrator construction and test (9Y) | | | | | | | | | | |
| **Resources** | 2026 | 2027 | 2028 | 2029 | 2030 | 2031 | 2032 | 2033 | 2034 | 2035 |
| Staff | 1.2 | 1.2 | 1.8 | 1.8 | 1.8 | 2.1 | 2.1 | 3.5 | 2.5 | |
| Postdoc/GRAD | 1.6 | 1.6 | 2.4 | 2.4 | 2.4 | 2.8 | 2.8 | 2.1 | 1.5 | |
| Student | 1.2 | 1.2 | 1.8 | 1.8 | 1.8 | 2.1 | 2.1 | 1.4 | 1 | |
| Material | 300 | 300 | 500 | 500 | 500 | 750 | 750 | 1000 | 1000 | |
| **Interested partners** | | | | | | | | | | |
| Academia: CERN, INFN, PSI, CEA, University of Twente, University of Southampton, Technical University Tampere | | | | | | | | | | |
| Industry: Tape manufacturers | | | | | | | | | | |

Table AI.4. Summary of objectives, main deliverables, personnel and material needs for the RCS magnet string and power system (RCS-String).

| Objectives | | | | | | | | | | |
|---|---|---|---|---|---|---|---|---|---|---|
| Build and test a string of resistive pulsed dipoles, including powering system and capacitor-based energy storage, aiming at field swing of +/- 1.8 T, maximum ramp-rate of 3.3 kT/s, and energy recovery efficiency better than 99 % | | | | | | | | | | |
| **High-level Deliverables** | | | | | | | | | | |
| 1) Dipole magnet stand-alone test (3Y) | | | | | | | | | | |
| 2) Power converter and energy storage stand-alone test (3Y) | | | | | | | | | | |
| 3) String construction and test (7Y) | | | | | | | | | | |
| **Resources** | 2026 | 2027 | 2028 | 2029 | 2030 | 2031 | 2032 | 2033 | 2034 | 2035 |
| Staff | 1.4 | 1.4 | 2.8 | 3.6 | 3.6 | 3 | 1 | | | |
| Postdoc/GRAD | 0.4 | 0.4 | 0.8 | 0.4 | 0.4 | 0 | 0 | | | |
| Student | 0.2 | 0.2 | 0.4 | 0 | 0 | 0 | 0 | | | |
| Material | 250 | 300 | 950 | 1500 | 1500 | 1000 | 500 | | | |
| **Interested partners** | | | | | | | | | | |
| Academia: CERN, University of Bologna, Technical University of Darmstadt | | | | | | | | | | |
| Industry: | | | | | | | | | | |

Table AI.5. Summary of objectives, main deliverables, personnel and material needs for the Wide-aperture, steady state Nb3Sn dipole (MBHY).

| Objectives | | | | | | | | | | |
|---|---|---|---|---|---|---|---|---|---|---|
| Build and test Nb3Sn demonstrator dipole with field target of 11 T, large bore, target 160 mm, 5 m long, operating with forced-flow of helium at 4.5 K | | | | | | | | | | |
| **High-level Deliverables** | | | | | | | | | | |
| 1) Dipole magnet engineering and validation tests (demonstrators) completed (5Y) | | | | | | | | | | |
| 2) Short model construction and test (9Y) | | | | | | | | | | |
| 3) *Magnet long prototype construction and test (11Y)* | | | | | | | | | | |
| **Resources** | 2026 | 2027 | 2028 | 2029 | 2030 | 2031 | 2032 | 2033 | 2034 | 2035 |
| Staff | 2 | 2 | 3 | 3.5 | 3.5 | 3.5 | 6.3 | 6.3 | 6.3 | 6.3 |
| Postdoc/GRAD | 1.2 | 1.2 | 1.8 | 2.1 | 2.1 | 2.1 | 1.8 | 1.8 | 1.8 | 1.8 |
| Student | 0.8 | 0.8 | 1.2 | 1.4 | 1.4 | 1.4 | 0.9 | 0.9 | 0.9 | 0.9 |
| Material | 300 | 500 | 750 | 845 | 750 | 750 | 1750 | 2000 | 2000 | 1500 |
| **Interested partners** | | | | | | | | | | |
| Academia: CERN, INFN | | | | | | | | | | |
| Industry: Nb3Sn manufacturers | | | | | | | | | | |

Table AI.6. Summary of objectives, main deliverables, personnel and material needs for the Rectangular aperture HTS dipole (MBHTS).

| Objectives | | | | | | | | | | |
|---|---|---|---|---|---|---|---|---|---|---|
| Build and test a 1 m long demonstrator for a HTS, 10 T, 30x100 mm bore dipole operating at 20 K | | | | | | | | | | |
| **High-level Deliverables** | | | | | | | | | | |
| 1) Dipole magnet engineering and validation tests (demonstrator) completed (5Y) | | | | | | | | | | |
| 2) Model construction and test (10Y) | | | | | | | | | | |
| **Resources** | 2026 | 2027 | 2028 | 2029 | 2030 | 2031 | 2032 | 2033 | 2034 | 2035 |
| Staff | 1.6 | 1.6 | 2.4 | 2.4 | 2.4 | 3.2 | 4 | 3.5 | 4.9 | 2.8 |
| Postdoc/GRAD | 1.2 | 1.2 | 1.8 | 1.8 | 1.8 | 2.4 | 2.4 | 2.1 | 1.4 | 0.8 |
| Student | 1.2 | 1.2 | 1.8 | 1.8 | 1.8 | 2.4 | 1.6 | 1.4 | 0.7 | 0.4 |
| Material | 200 | 200 | 500 | 500 | 850 | 1500 | 1500 | 1250 | 1250 | 500 |
| **Interested partners** | | | | | | | | | | |
| Academia: CERN, INFN, Technical University Tampere | | | | | | | | | | |
| Industry: Tape manufacturers | | | | | | | | | | |

Table AI.7. Summary of objectives, main deliverables, personnel and material needs for the Wide aperture HTS dipole (MBHTSY).

| Objectives | | | | | | | | | | |
|---|---|---|---|---|---|---|---|---|---|---|
| Build and test a 1 m long demonstrator for a HTS, 14 T, 140 mm bore dipole operating at 20 K | | | | | | | | | | |
| **High-level Deliverables** | | | | | | | | | | |
| 1) Dipole magnet engineering and validation tests (demonstrator) completed (6Y) | | | | | | | | | | |
| 2) Short model construction and test (16Y) | | | | | | | | | | |
| 3) Long prototype construction and test beyond the scope of this proposal (20Y) | | | | | | | | | | |
| **Resources** | 2026 | 2027 | 2028 | 2029 | 2030 | 2031 | 2032 | 2033 | 2034 | 2035 |
| Staff | 2 | 2.4 | 2.4 | 3.2 | 4 | 4 | 4 | 4 | 4.5 | 6.3 |
| Postdoc/GRAD | 1.5 | 1.8 | 1.8 | 2.4 | 2.4 | 2.4 | 2.4 | 2.4 | 2.7 | 1.8 |
| Student | 1.5 | 1.8 | 1.8 | 2.4 | 1.6 | 1.6 | 1.6 | 1.6 | 1.8 | 0.9 |
| Material | 300 | 500 | 750 | 800 | 800 | 800 | 800 | 800 | 1100 | 1250 |
| **Interested partners** | | | | | | | | | | |
| Academia: CERN, INFN, Technical University Tampere | | | | | | | | | | |
| Industry: Tape manufacturers | | | | | | | | | | |

Table AI.8. Summary of objectives, main deliverables, personnel and material needs for the Wide aperture HTS IR quadrupole (MQHTSY).

| Objectives | | | | | | | | | | |
|---|---|---|---|---|---|---|---|---|---|---|
| Build and test a 1 m long demonstrator for a HTS, 300 T/m, 140 mm bore quadrupole operating at 20 K | | | | | | | | | | |
| **High-level Deliverables** | | | | | | | | | | |
| 1) Quadrupole magnet engineering and first validation tests (demonstrator) completed (7Y) | | | | | | | | | | |
| 2) Short model construction and test beyond the scope of this proposal (16Y) | | | | | | | | | | |
| 3) Long prototype construction and test beyond the scope of this proposal (20Y) | | | | | | | | | | |
| **Resources** | 2026 | 2027 | 2028 | 2029 | 2030 | 2031 | 2032 | 2033 | 2034 | 2035 |
| Staff | 0 | 0 | 0 | 0 | 1.5 | 2 | 2 | 2 | 3 | 4.2 |
| Postdoc/GRAD | 0 | 0 | 0 | 0 | 0.9 | 1.2 | 1.2 | 1.2 | 1.8 | 1.2 |
| Student | 0 | 0 | 0 | 0 | 0.6 | 0.8 | 0.8 | 0.8 | 1.2 | 0.6 |
| Material | 0 | 0 | 0 | 0 | 200 | 200 | 500 | 750 | 850 | 1000 |
| **Interested partners** | | | | | | | | | | |
| Academia: CERN, INFN, Technical University Tampere | | | | | | | | | | |
| Industry: Tape manufacturers | | | | | | | | | | |

Table AI.9. Summary of objectives, main deliverables, personnel and material needs for the R&D on materials and methods.

| Objectives | | | | | | | | | | |
|---|---|---|---|---|---|---|---|---|---|---|
| Host and coordinate methods and materials R&D, characterization and testing common to magnet demonstrators design, manufacturing and testing | | | | | | | | | | |
| **High-level Deliverables** | | | | | | | | | | |
| 1) HTS magnets design code (5Y) | | | | | | | | | | |
| **Resources** | 2026 | 2027 | 2028 | 2029 | 2030 | 2031 | 2032 | 2033 | 2034 | 2035 |
| Staff | 1.2 | 1.2 | 1.2 | 1.2 | 1.2 | 1.2 | 2.1 | 2.1 | 2.1 | 2.1 |
| Postdoc/GRAD | 0.9 | 0.9 | 0.9 | 0.9 | 0.9 | 0.9 | 0.6 | 0.6 | 0.6 | 0.6 |
| Student | 0.9 | 0.9 | 0.9 | 0.9 | 0.9 | 0.9 | 0.3 | 0.3 | 0.3 | 0.3 |
| Material | 200 | 200 | 200 | 200 | 200 | 400 | 400 | 400 | 400 | 400 |
| **Interested partners** | | | | | | | | | | |
| Academia: CERN, INFN, University of Twente, University of Southampton, Technical University Tampere, KEK | | | | | | | | | | |
| Industry: | | | | | | | | | | |

**Appendix II. Definition of Technology Readiness Level**

Quoting the definition from [29]:

"*Technology readiness levels (TRLs) are a method for estimating the maturity of technologies during the acquisition phase of a program [...] TRL is determined during a technology readiness assessment (TRA) that examines program concepts, technology requirements, and demonstrated technology capabilities. TRLs are based on a scale from 1 to 9 with 9 being the most mature technology.*"

For reference, we report below the scale of TRL and definitions adopted by the EU, based on the original NASA formulation.

| TRL | EU Definition |
|---|---|
| 1 | Basic principles observed |
| 2 | Technology concept formulated |
| 3 | Experimental proof of concept |
| 4 | Technology validated in lab |
| 5 | Technology validated in relevant environment (industrially relevant environment in the case of key enabling technologies) |
| 6 | Technology demonstrated in relevant environment (industrially relevant environment in the case of key enabling technologies) |
| 7 | System prototype demonstration in operational environment |
| 8 | System complete and qualified |
| 9 | Actual system proven in operational environment (competitive manufacturing in the case of key enabling technologies; or in space) |